\documentclass[12pt]{article}
\def\lsim{\mathrel{\rlap {\raise.5ex\hbox{$ < $}}
{\lower.5ex\hbox{$\sim$}}}}
\def\gsim{\mathrel{\rlap {\raise.5ex\hbox{$ > $}}
{\lower.5ex\hbox{$\sim$}}}}
\newcommand{\ba}{\begin{eqnarray}}
\newcommand{\ea}{\end{eqnarray}}
\newcommand{\be}{\begin{equation}}
\newcommand{\ee}{\end{equation}}
\def\np#1#2#3{{\it {Nucl. Phys.}} {\bf{B#1}} (#2) #3}
\def\pl#1#2#3{{\it {Phys. Lett.}} {\bf{B#1}} (#2) #3}
\def\prl#1#2#3{{\it {Phys. Rev. Lett. }}{\bf{#1}} (#2) #3}
\def\pr#1#2#3{{\it {Phys. Rev.}} {\bf{D#1}} (#2) #3}

\begin{document}
\begin{titlepage}
\begin{flushright}

hep-ph/9807289{\hskip.5cm}\\  
\end{flushright}
\begin{centering}
\vspace{.3in}
{\bf SUPPRESSING DIMENSION FIVE OPERATORS IN GENERAL $SU(5)$ MODELS
$^\ast$ }\\
\vspace{2 cm}
{M. E.  G\'OMEZ, J. RIZOS
 and K. TAMVAKIS}\\
\vskip 1cm

{\it {Physics Department, University of Ioannina\\
Ioannina, GR45110, GREECE}}\\

\vspace{1.5cm}
{\bf Abstract}\\
\end{centering}
\vspace{.1in}
We discuss dimension--5 operators in supersymmetric models 
containing 
extra 
hypercharge--$1/3$ colour--triplets. We derive a 
general  formula relating dimension--5 operator to the colour--triplet mass matrix.
We show that certain zeros in the triplet mass--matrix together with 
some triplet coupling selection rules can lead to elimination of 
dimension--5 operators.
In particular we focus on $SU(5)$ models and 
show that \\
(a) Dimension--5 operators can be 
eliminated in the standard $SU(5)$ model by the introduction of an extra pair 
of ${\bf5}+\overline{\bf5}$ Higgses with specific couplings\\
(b) Flipped $SU(5)$ models
with extra ${\bf{10}}+\overline{{{\bf10}}}$ Higgses are free of dimension--5 operators\\
(c) Flipped $SU(5)$ models with extra ${\bf{5}}+\overline{\bf5}$ and/or extra 
${\bf{10}}+\overline{\bf10}$ Higgses can be made free of dimension--5 operators  
for a textured form of the triplet mass--matrix accompanied by 
constraints on the $5$--plet couplings to matter.\\
Our analysis is motivated by the recently put forward M--theory phenomenological 
framework that requires a strong string coupling and reintroduces the problem of eliminating 
dimension--5 operators.

\vspace{1cm}
\begin{flushleft} 
July 1998
\end{flushleft}
\hrule width 6.7cm \vskip.1mm{\small \small
$^\ast$ Research partially supported by the EEC under the contract 
ERBFMRX-CT96-0090.}

 \end{titlepage}
\section{Introduction}

The Standard Model and its $N=1$ supersymmetric extension\cite{NHK},
the Minimal Supersymmetric 
Standard Model (MSSM), can be naturally embedded in a Grand Unified Theory
\cite{GUTS} (GUT) 
with interesting phenomenological 
and cosmological consequences. GUTs can successfully predict the electroweak 
mixing angle $\sin^2\theta_W$ \cite{USSM}, fermion 
mass relations as well as provide a mechanism for the creation of the baryon
asymmetry of the Universe\cite{BA}. A GUT can be in principle 
accommodated in the 
framework of Superstring Theory\cite{GSW}.

Proton decay is 
a generic feature of any unification scheme since the unification of quarks and 
leptons in a common multiplet introduces extra interactions that violate baryon number. 
Proton decay rates and modes are a prediction of GUT models that play a crucial role in 
their phenomenological viability. In fact, proton decay has turned out to be the nemesis of 
many GUT and Superstring models. It is a welcome prediction that can be used 
to test GUTs. In supersymmetric GUTs with conserved $R$--parity the dominant 
baryon number violating operators are dimension $D=5$, while $D=6$ operators are in general suppressed due 
to the increase of the unification scale in comparison to its non--supersymmetric values. $D=5$ 
 operators are proportional to the Yukawa couplings and to the inverse 
of the  heavy mass \cite{DIMF}. In minimal models the 
Yukawa couplings involved are associated with the fermion masses. The values 
of these couplings play an important role in the final value of the proton 
decay rate and the resulting hierarchy of existing modes. 
Nevertheless, Superstring embeddable models \cite{SEM} or 
models of phenomenologically oriented GUTs that treat the fermion mass problem
 \cite{FMM}, come out with an extended Higgs sector. 
Similarly, several attempts for the construction of 
phenomenologically viable string models have yielded up to now models 
with extra Higgses \cite{SM}. 
Motivation for an extended Higgs sector can also 
be found in fixed--point considerations \cite{FP}.
In addition, the presence of extra hypercharge $1/3$ colour--triplets 
turns out to be a  necessary ingredient of models that raise
the GUT unification scale close to the string scale \cite{AEN,ET}.

The Yukawa couplings of Higgs fields to matter multiplets are 
constrained phenomenologically 
by the observed quark and lepton mass--matrices. Nevertheless, in the presence of more than one pair 
of Higgses not all coupling constants are constrained. In unified models these same Yukawa coupling 
constants determine the strength of Baryon number violating interactions. Therefore, not all Yukawas 
appearing in front of dimension--5 Baryon number violating operators are necessarily small from 
fermion mass considerations. Since the proton decay rate through such operators
$${\cal{O}}(\frac{\alpha}{16\pi})(\frac{Y^2}{m_{\tilde{H}}m_S})(QQQL)$$
 is predicted to be 
 not very far away from the experimental limit, the presence of large, i.e. not as small as the fermion mass 
 matrix requires, Yukawas could be disastrous. Very roughly, the imposed constraint is 
 $Y\leq 10^{-6}/M_P$ . Instead of extending the fermion mass constraint to all 
 Yukawas one could investigate other ways, such as symmetries or 
 selection rules imposed by the Higgs mass matrix, 
that could lead to the elimination or suppression of the 
 related dangerous dimension--5 operators.
Various ideas have been reported to the literature in order to sufficiently 
suppress such operators \cite{DMS}.

Additional motivation comes from the recent theoretical developments have put 
forward a version of $M$--theory which 
 yields the {\it{strong coupling }}limit of the $E_8\times E_8$ heterotic string \cite{XYZ}. This formulation involves 
 a 11--dimensional ``bulk" space with 10-dimensional ``walls" at each of each ends. This construction 
 circumvents the discrepancy between the string unification scale and the grand unification scale. 
 In an ``M--phenomenology" framework a generic non--renormalizable interaction that violates Baryon number would be of the form
 ${\cal{O}}(1){\frac{g^{N+1}_s}{M^{N+1}}}{\Phi^N}(QQQL)$
 where $M\sim 10^{16}$ GeV, $\Phi/M\sim {\cal{O}}(1)$ and 
$g_s\sim {\cal{O}}(1)$
\cite{ELLIS}. 
Therefore, it is clear that in 
 such a framework dimension--5 operators have to be eliminated in 
 a drastic way since the standard suppression argument is not sufficient.
Of course non--renormalizable interactions are not the only source of 
dimension--5 operators in string derived models.
As in the case of GUTs an effective dimension--5 operator can be formed by 
gluing together renormalizable interactions of the effective langangian and this
class of operators needs also to be suppressed as it will  involve some strong 
Yukawa couplings.

In this paper we examine  dimension--5 operators 
in the context of general GUT models containing extra hypercharge 1/3 colour
triplets and particularly in  $SU(5)$ models.  We propose a mechanism 
for eliminating or suppressing such operators based on the use of   
textures of the hypercharge 1/3 mass--matrix accompanied by certain 
constraints of the extra triplet coupling to matter.
 
\section{Dimension--5 operators in models with extra D--quark triplets}
 Let us consider a general supersymmetric model containing some extra hypercharge--$1/3$ colour--triplets
\footnote{This superpotential arises in the case of the standard $SU(5)$ with extra
 Higgs 5--plets or from the 
flipped $SU(5)\times U(1)$ with both extra Higgs 5--plets and 10--plets. 
However the results of Sections 1 and 2 apply to the wider class of GUTs that share
the same effective superpotential.}.
The effective $SU(3)_c\times SU(2)_L\times U(1)_Y$ superpotential describing the 
couplings of quarks and leptons to the extra coloured triplets of the $D$--quark type
\begin{equation}
f_{ij}^{\alpha}Q_i Q_j D_{\alpha}+y_{ij}^{\alpha} Q_i L_j{\overline{D}}_{\alpha}
 + r_{ij}^{\alpha}E_i^{c}U_j^{c}{\overline{D}}_{\alpha} \ ,
\label{ww}
\end{equation}
where $i,j=1,2,3$ are the usual generation indices and $\alpha=1,...,N$ is an extra 
index describing the multiplicity of triplets
and repeated indices are summed. In addition the effective triplet mass matrix
will be of the form
\begin{equation}
{({\cal M}_3)}_{\alpha \beta}D_{\alpha}\overline{D}_{\beta}
\end{equation}
where ${\cal M}_3$ is in general non--diagonal.  
 
We can always go to a basis in which the triplet mass--matrix is diagonal
\begin{equation}
D_{\alpha}=S_{\alpha \beta}D_{\beta}^\prime\,\,\,\,,\,\,\,
\overline{D}_{\alpha} = U_{\alpha \beta}\overline{D}_{\beta}^\prime
\end{equation}
\begin{equation}
{{\cal M}_3}_D\equiv diag (m_1,m_2,\cdots,m_N)=S^{T}{{\cal M}_3}\,U
\end{equation}
where the matrices $S$ and $U$ are unitary.  In this basis we can easily evaluate
 $D=5$ operators resulting from Higgs triplet fermion exchange, and then recast the 
result in the original basis.
Assuming that all triplets are massive ($m_i\ne0\,,i=1,\dots, N$),
operators with the structure $Q_i\,Q_j\,Q_k\,L_n$ will be proportional to
\footnote{The corresponding four--fermion operator, assuming roughly 
an overall universal 
supersymmetry breaking scale  $m_S$ , will involve
 ${\frac{({\cal{M}}_3)^{-1}}{m_S}}
-4 m_S({{\cal{M}}_3})^{-3}
\log{\frac{({{\cal{M}}_{3}})}{m_S}}$ .} 
\ba
{\cal O}^{\mbox{\it\tiny QQQL}}_{ijkl}&=&
\sum_{\alpha,\beta,\gamma=1}^{N}f_{ij}^{\alpha}
S_{\alpha\gamma}({{\cal M}_3}_{D}^{-1})_{\gamma}U_{\beta\gamma}y_{kn}^{\beta}\nonumber\\
&=&\sum_{\alpha ,\beta =1}^{N}f_{ij}^{\alpha}({{\cal M}_3}^{-1})_{\alpha \beta}^{T}
y_{kn}^{\beta}=\frac{1}{\det({{\cal M}_3})}\,\sum_{\alpha,\beta=1}^{N}
f_{ij}^{\alpha}\,{\rm cof}({{\cal M}_3})_{\alpha\beta}\,y_{kn}^{\beta}
\label{ma}
\ea
Analogous formulas hold for $D=5$ operators of the type $Q_iQ_jU^c_kE^c_k$.

\section{How to suppress dimension--5 operators in effective models with extra triplets}

Suppose now that we want to eliminate all dimension five operators. 
Assuming that 
the Yukawa couplings $f_{ij}^\alpha$ and $y_{ij}^\beta$ are in general unrelated  and
$\det {\cal M}_3\ne0\,$, equation (\ref{ma}) implies that 
{\em 
the necessary and
sufficient condition for vanishing of the ${\cal O}^{\mbox{\it\tiny QQQL}}_{ijkl}$ operator is that
for every pair of triplets ($D^\alpha$,${\overline{D}}^\beta$\ ,
$\alpha,\beta=1,\dots,N$) that do couple to quarks and leptons 
 respectively ($f^\alpha_{ij}\ne0\ and\ h^\beta_{ij}\ne0$)
the cofactor of the corresponding triplet mass matrix element $({\cal M}_3)_{\alpha\beta}$
vanishes}\footnote{We consider here the triplet $D^\alpha$ as coupled 
to quarks and leptons if at least one 
 $f^\alpha_{ij}\ne0$ and similarly for anti--triplets.}
\be
{\cal O}^{\mbox{\it\tiny QQQL}}_{ijkl}=0 \Longleftrightarrow
{\rm cof}({\cal M}_3)_{\alpha\beta}=0\ \forall\ (\alpha,\beta)\in 
\Xi=\{(\alpha,\beta) : f^\alpha_{ij}\ne0\ {\rm and}\  h^\beta_{kl}\ne0\}\ .
\label{con}
\ee
It is obvious that in the case where all triplets ($D$'s and $\overline{D}$'s) 
couple to matter
the suppression of dimension five operators (\ref{ma}) is not possible since
(\ref{con}) leads to 
$\det(M_3)=0$.
Nevertheless, if for some reason (discrete symmetry, R--parity, anomalous $U(1)$, 
accidental symmetry) some of the $f^\alpha_{ij}$ and/or $y^\beta_{kl}$  are zero
and the triplet mass matrix is such that the cofactors of the appropriate 
 matrix elements are zero then the associated dimension--5 operator 
vanishes. 

The previous discussion leads to the possibility that {\em in a model with extra D--quark triplets
dimension--5 operators can be  eliminated by using  textures of triplet mass matrices
and the triplet--matter couplings.}

To be concrete let us give a simple example of such an effective theory. Consider 
the case of an effective theory with two extra triplets. Only the first 
couples  to matter through the superpotential terms
\begin{equation}
 f_{ij}^{1}Q_iQ_jD_{1}+y_{ij}^{1}Q_iL_j{\overline{D}}_{1}
+r_{ij}^{1}E_i^{c}U_j^{c}{\overline{D}}_{1}
\label{tsp}
\end{equation}
and their mass--matrix has the form
\be
{\cal M}_3=\left(\begin{array}{cc}\mu_{11}&\mu_{12}\\ \mu_{21}&0\end{array}\right)\ .
\label{tmm}
\ee
Since $f_{ij}^{2}=y_{ij}^{2}=0\,$ evaluation of (\ref{ma}) leads to
\be
{\cal O}^{\mbox{\it\tiny QQQL}}_{ijkl}= 
f_{ij}^{1}\,{\rm cof}({{\cal M}_3})_{11}\,y_{kn}^{1} \sim {\rm cof}({{\cal M}_3})_{11} =0
\ee
It is remarkable that if we remove the second pair of triplets of the model
(which do not couple directly to matter) the usual dimension--5 operators 
reappear.
We shall study below that this nice property can be incorporated 
 in $SU(5)$ models.

Trying to generalize the previous model, let us
 consider an effective model that contains 
 $N_c$ triplets that directly couple to matter and
$N_u$ triplets that do not couple directly to matter. The most
general superpotential will be still that of equation (\ref{ww})
while the triplet mass matrix will be of the form
\begin{equation} 
{\cal{M}}_3=\left(\begin{array}{cc}
{\mu}_{\alpha\beta}&v_{\alpha A}\\
{\overline{v}}_{A\beta}&m_{AB}
\end{array}\right)
\end{equation}
where
 $A, B=1,\cdots,N_{u}$ ,  $\alpha, \beta, \gamma=1,\cdots,N_c\,$ and we 
assume that all triplets are heavy,i.e. $\det{\cal M}_3\ne0$.

For simplicity let us first consider the case $N_c=1$. In that case 
only one matrix element gives rise to proton decay and it is proportional
to
\be
{\rm cof}({\cal M}_3)_{11}=\frac{\det m}{\det{\cal M}_3}
\ee
which means that proton decay is absent only in the case that the restricted
  mass--matrix of the triplets not coupled to matter has 
\be
\det m =0
\label{cc}
\ee
We will see in Section \ref{secf} how this constrain naturally arises in the
context of the flipped $SU(5)\times U(1)$ model.

We can now generalize this idea by considering a model with $N_c$ and 
$N_u$ arbitrary. In this case we can prove
that the $N_{c}\times N_{c} $ submatrix of ${\cal M}_{3}^{-1}$ involved 
in the generation of $D=5$ operators will  vanish in 
theories with $ {N_{u}\geq N_{c} }$ when 
the matrix $m$ has a number $ N_{c}$ of null eigenvalues.

 Although we do not
give here a formal  proof of the last statement, it is easy to illustrate 
it by considering the case of 
 a model with $N_{c}= N_{u} $. 
$D=5$ operators can be eliminated when $m$ is a null matrix, which implies
\begin{equation} 
{\cal{M}}_3^{-1}=\left(\begin{array}{cc}
0 &{\overline{v}}^{-1}\\
v^{-1}&-v^{-1} \mu {\overline{v}}^{-1} 
\end{array}\right)
\end{equation}

	In effective theories with  $N_{c}>N_{u}$ it is not possible to make  all
the cofactors of the triplets coupled to matter vanish while keeping 
${\rm det}\left({\cal{M}}_3\right) \neq 0$. In effective theories with $N_{u}>N_{c}$ a number
$N_{c}$ of null eigenvalues of the submatrix $m$ (the mass matrix 
of the triplets that 
do not couple directly to matter) is required in order to  eliminate $D=5$ operators, while all the 
triplets remain heavy.

\section{An $SU(5)$ model without dimension--5 operators}
 Supersymmetric $SU(5)$ models predict proton decay through $D=5$ operators.
The predicted rates for the dominant mode $p\rightarrow \overline{\nu}\  K^{+}$
 are in the range of the experimental searches \cite{NA1}. It is interesting to 
analyze how this channel can be suppressed in models with several triplets.
 
Let us consider now an $SU(5)$ model with two pairs of Higgs pentaplets
${h}_\alpha,{\overline{h}}_\alpha\,, \alpha=1,2$ of which only the 
first couples to matter.
The quarks and leptons are  assigned to 
$\phi({\bf\overline{5}})+\psi({\bf\overline{10}})$ representations of $SU(5)$.
The 
part of the superpotential related to dimension--5 decay will  be
\begin{equation}
f_{ij} \,\psi_i \psi_j {h}_{1} + y_{ij}\,\psi_i\phi_j
 {\overline{h}}_{1}+\sum_{\alpha,\beta=1}^2({\mu}_{\alpha \beta}{h}_{\alpha}
{\overline{h}}_{\beta}
+{\lambda}_{\alpha \beta}{h}_{\alpha}{\Sigma}{\overline{h}}_{\beta})
\ ,
\label{su5s}
\end{equation}
where the symbol  $\Sigma$ stands for 
the adjoint Higgs superfield in the {\bf{24}} representation.
The isodoublet and colour--triplet  mass matrices are correspondingly
 of the form
\begin{equation}
{\cal{M}}_2=\mu-3{\lambda}V \ ,
\end{equation}
\begin{equation}{\cal{M}}_3=\mu+2{\lambda}V\end{equation}
The well known fine--tuning that guarantees a massless pair of isodoublets amounts to 
\begin{equation}
\det({\cal{M}}_2)=0 \ .
\label{ft}
\end{equation}
The proton decay rate through $D=5$ operators is, according to equation (\ref{ma}),
 determined 
by the cofactor of the $1-1$ element  of the  triplet mass matrix 
\begin{equation}
{\rm cof}({\cal M}_3)_{11}={(\mu_{22}+2{\lambda}_{22}V)} \ .
\end{equation}
Hence, choosing $\mu_{22}=-2{\lambda}_{22}V$ dimension--5 operators vanish. 
This condition  is perfectly compatible with the previous fine--tuning condition (\ref{ft}). 
It is very interesting that 
proton decay through $D=5$ operators can be set to zero through a condition on the couplings 
\footnote{
Of course, proton decay still goes through at the (suppressed) rate of $D=6$ operators.}.

In the framework of our standard $SU(5)$ example the required zero in the 
inverse triplet mass 
matrix does not correspond to any symmetry and is in a sense a 
second fine--tuning \cite{NA2}.
 Nevertheless, the general conclusion is that zeros of the triplet mass matrix,
 perhaps attributable to symmetries, can stabilize the proton. 

The superpotential considered above in (\ref{su5s}) is not the most general one.
In fact,  the case that all 5--plets couple to matter cannot be reduced
to (\ref{su5s}) since it would require a different Higgs 5--plet rotation 
for each generation of matter. Therefore, as shown in Section 3,
 in order to suppress dimension--5 
operators in models with more that one ($N_c$) 5--plets 
coupled to matter  must have an equal or bigger number ($N_u$) of uncoupled 
5--plets. In addition the mass matrix of the uncoupled 5--plets should 
have a number of $N_c$ null eigenvalues. This means we need
in general $N_c$ fine--tunings in addition to the one related to the 
doublet--matrix.

The suppression of $D=5$ operators by heavy triplet masses, as it is required in
the minimal $SU(5)$ is very restricted \cite{KS}. Therefore constraints imposed
by proton decay rule out most of the parameter space for this model 
\cite{NA2}. We should emphasize the fact that in $SU(5)$ models with non minimal 
Higgs content, the constraints imposed by proton decay on the parameter space 
and triplet masses can be relaxed.

\section{\label{secf}Dimension--5 operators in extensions of the flipped $SU(5)$ }
 In the minimal flipped $SU(5)\times U(1)$ 
model \cite{BN} matter fields come in the representations 
\begin{equation}
{F}_i({\bf{10}}, 1/2)\,\,\,,\,\,\,\,\,{f}_i^c({\bf{\overline{5}}},-3/2)\,\,,
\,\,\,\,\,{l}_i^c({\bf{1}},5/2)
\end{equation}
while Higgses in 
\begin{equation}
{h}({\bf{5}},-1)\,\,\,,\,\,\,{\overline{h}}
({\overline{\bf{5}}}, 1)\ ,
\end{equation}
and in 
\begin{equation}
{F}_{h}({\bf{10}}, 1/2)\,\,\,,\,\,\,\,\,{\overline{F}}_h
({\overline{\bf{10}}},-1/2)\ .
\end{equation}

A great advantage of the  ``flipped" $SU(5)$ model over the ordinary one is 
that of the realization of the ``triplet--doublet splitting" mechanism 
through which 
the Higgs isodoublets that are about to achieve the electroweak breaking 
stay massless 
while the colour triplet Higgses obtain a large mass. 
This mechanism
allows us  to  naturally suppress dimension--6 operators  but not
 the dimension--5 ones. 
The suppression of dimension--5 operators in the  minimal $SU(5)\times U(1)$ model is 
due 
to the triplet mass matrix texture and it is a subcase of 
the mechanism described in Section 1.
It is crucial that R--parity does not allow the Higgs
10-plets to mix with matter and the effective superpotential has the 
form (\ref{tsp}). We denote $D_1, \overline{D}_1$ the triplet pair 
that lies in the 
$({\bf5},{\bf1})+(\overline{\bf5},{\bf1})$ reps.  The triplet mass matrix has the form
(\ref{tmm}) (with $\mu_{11}=0$ in the minimal model). Another nice feature 
here that should be stressed is that $\mu_{22}=0$ as dictated by F--flatness.
The pair $F_h\, {\overline{F}_h} $ has to be massless in order to realize the
$SU(5)\times U(1)$ breaking to the standard model.

In spite of the nice features of the minimal flipped $SU(5)$ model, 
all attempts to obtain such a model from strings have yielded up to now
non--minimal models. Such models include\\
(a) extra pairs of low energy Higges ($h$, $\bar h$) and/or\\
(b) extra pairs of $SU(5)\times U(1)$ breaking Higges ($F_h$, ${\overline{F}_h}$).

We are thus motivated to study the presence of dimension--5 operators in
such models. 
As we shall see contrary to the minimal case, such extensions
of the flipped $SU(5)$ model are not automatically free of dimension--5 
operators. 

The relevant part of the superpotential assuming $N_5$ pairs of Higgs 
5--plets 
($h_\alpha, {\overline{h}_\alpha}\ , \alpha=1,\dots,N_5$) that couple to matter and
$N_{10}$ pairs of Higgs 10--plets
($F_{h\alpha}, {\overline{F}_{hA}}\ , A=1,\dots,N_{10}$) that do not couple to matter, will
have the form
\begin{eqnarray}
&&
f_{ij}^{\alpha}F_i\,F_j\,{h}_{\alpha}+y_{ij}^{\alpha}F_i\,f_j^{c}
{\overline{h}}_{\alpha}+r_{ij}^{\alpha}l_i^{c}\,f_j^{c}\,{\overline{h}}_{\alpha}+
{\mu_{\alpha\beta}}{h}_{\alpha} {\overline{h}}_{\beta}+
m_{AB}{F}_{hA}{\overline{F}}_{hB}\nonumber\\
&&\ \ \ \ \ \ \ \ \ \ \ \ \ \ \ \ \ \ \ \ \  \ \ \ \ \ \ \ \ \ \ \ 
\mbox{}+{\lambda}_{AB\gamma}
{F}_{hA}{F}_{hB}{h}_{\gamma}
+{\overline{\lambda}}_{AB\gamma}
{\overline{F}}_{hA}{\overline{F}}_{hB}{\overline{h}}_{\gamma}
\end{eqnarray}
where $A, B=1,\cdots,N_{10}$ ,  $\alpha, \beta, \gamma=1,\cdots,N_5\,$.
Assuming GUT symmetry breaking to an arbitrary direction in the Higgs 
$10$--plet space ($(V_1,V_2,\dots,V_{N_{10}})$ and similarly for bars)
\footnote{D--flatness requires $\sum_A V_A^2=\sum_A{\overline{V}_A}^2$}, 
we obtain the triplet mass matrix
\footnote{In a $(D_1,\cdots,D_{N_5}, (d_H^c)_1,\cdots,(d_H^c)_{N_{10}})$ 
versus $({\overline{D}}_1,\cdots,
{\overline{D}}_{N_5}, ({\overline{d}}_H^c)_1,\cdots, 
({\overline{d}}_H^c)_{N_{10}})$ basis, where with  $D$ we denote
the triplets which lie inside the Higgs 5--plets and with $D_{H}$ the triplets
that lie inside the Higgs 10--plets. }
\begin{equation} 
{\cal{M}}_3=\left(\begin{array}{cc}
{\mu}_{\alpha\beta}&v_{\alpha A}\\
{\overline{v}}_{A\beta}&m_{AB}
\end{array}\right)
\end{equation}
where $\mu_{\alpha\beta}$ is the doublet mass--matrix and
 $v_{\alpha A}=2{\lambda_{AB\alpha}} V_B\,$,
${\overline{v}}_{A\beta}=2{\overline{\lambda}}_{AB\beta} {\overline{V}}_B\,$

Let us now start our study by a simple example.
Consider the flipped model with two pairs of Higgs 5--plets and one pair of 
Higgs 10--plets.
Assuming for simplicity that the 5-plet mass matrix is diagonal,
the explicit form of the triplet matrix is 
\footnote{We have renamed $\lambda_1=\lambda_{111}\,, \lambda_2=\lambda_{112}$.}
\begin{equation}
{\cal{M}}_3=\left(\begin{array}{ccc}
0&0&{\lambda}_1V\\
0&\mu&{\lambda}_2V\\
{\overline{\lambda}}_1{\overline{V}}&{\overline{\lambda}}_2{\overline{V}}&0
\end{array}\right)
\end{equation}
and $\det({\cal M}_3)=\lambda_1{\overline{\lambda}}_1{\overline{V}}$
The transpose of inverse triplet matrix entering in formula (\ref{ma}) is 
\begin{equation}
\left({{\cal{M}}_3^{-1}}\right)^T= \left(
\begin{array}{ccc}
{\frac{\lambda_2{\overline{\lambda}}_2}{\lambda_1{\overline{\lambda}}_1\mu}}&
-\frac{{{\lambda}}_2} {{{\lambda}}_1\mu}&
\cdot\\
-\frac{{\overline{\lambda}}_2}{{\overline{\lambda}}_1\mu}
&\frac{1}{\mu}&\cdot\\
\cdot&\cdot&\cdot
\end{array}\right)
\label{myma}
\end{equation}
where the dots  stand for elements which are irrelevant. It is now obvious that in
this  model dimension five operators cannot be eliminated since even in the
case $\overline{\lambda}_2=\lambda_2=0$ the $22$ element does not vanish.
If we want to eliminate them we have two solutions :\\
(i) assume that the extra pair of 5--plets does not couple to matter. In this
case only the $11$ element of the matrix in (\ref{myma}) is relevant and 
it vanishes for $\lambda_2=0$ (or ${\overline{\lambda}_2=0}$).\\
(ii) make the milder assumption that one of the 5--plets (e.g. $h_2$)
 does not couple to the up quarks (or similarly $\overline{h}_2$ does not couple
to the down).  In this case the second column (or line) of the matrix in 
(\ref{myma}) becomes irrelevant and  the column (or line) left vanishes for
 $\lambda_2=0$ (or $\overline{\lambda}_2=0$).

Another case that could arise is the existence of extra decuplets. The simplest of 
these cases is for $N_5=1$ and $N_{10}=n\geq 2$. This corresponds to the case
$N_c=1, N_u=n$ studied in Section 3.
The interesting feature here is that the requirement (\ref{cc}) 
is automatically satisfied. Since decoupled triplets have the same mass 
matrix as Higgs decuplets, this constraint arises as a consequence of 
F--flatness 
which demands ${\det(m)=0}$ in order to have at least one pair of massless 
Higgs decuplets which will realize the GUT symmetry breaking.
One can actually choose $m$ to have only one zero eigenvalue so that 
all remnants of the Higgs decuplets will become heavy.

In the more general case where $N_5$ and $N_{10}$ are arbitrary 
dimension--5 operators can be suppressed only in the case $N_5\le N_{10}$ 
according to our analysis in Section 3. Furthermore one has to require 
that the Higgs decuplet mass matrix has $N_5$ zero eigenvalues.
This is compatible with symmetry breaking and with the requirement of 
making all triplets heavy but leaves $N_5-1$ pairs of $Q({\bf3},{\bf2},1/6)+
{\overline Q}({\bf\bar3},{\bf2},-1/6)$ massless.
This feature does not necessarily mean that this possibility is ruled out.
On the contrary one can consider the cases where extra $Q$'s have intermediate 
masses which are small enough to sufficiently suppress dimension--5 operators
but they are still compatible with renormalization group requirements.
The appearance of extra vector--like pairs of $Q$ and $D$ type multiplets with intermediate 
masses is a welcomed feature in the context of flipped $SU(5)\times U(1)\,$ 
models that raise the unification scale to the string scale \cite{AEN,LN}. 

\section{Conclusions }
In the present article we considered the problem of dimension--5 operators that violate Baryon number. Models 
with extended Higgs sectors can in general have such operators and their strength cannot be assumed to be 
within the allowed limits since the standard Yukawa coupling suppression argument is in itself a new more
 or less ad hoc constraint. We, therefore, considered the problem of the elimination of these operators. An 
 additional motivation comes about from the recently proposed M--phenomenology framework which corresponds to 
 an effective theory with strong string coupling.

  We followed a phenomenological approach trying to be general and restricting 
 ourselves mainly on finding the necessary and sufficient conditions that can lead to the elimination of 
 dimension--5 operators. Some of these conditions could be shown to correspond to a particular symmetry. 
 Nevertheless, we did not insist on the point of symmetry since the theories under consideration are supposed to 
 be effective and the symmetry structure of their renormalizable part does not always coincide with the symmetry 
 structure of the complete theory. 

Our main result is that textured zeros of the color--triplet mass--matrix as well as 
 Yukawa selection rules can eliminate certain dimension--5 operators. In order to be specific we focused on $SU(5)$ 
 models. 
 In particular, we showed that introducing an extra pair of Higgs pentaplets in the standard supersymmetric $SU(5)\,$,
  with specific couplings, can eliminate 
 these operators. We also considered the case of the flipped--$SU(5)$ model with extra pentaplets and decuplets 
 and analyzed the conditions for vanishing proton decay through dimension--5 operators. Flipped--$SU(5)$ with 
 extra decuplets was shown to be $D=5$ operator--free as it happens in the
case of the minimal model.
However, flipped--$SU(5)$ with extra Higgs pentaplets is
not automatically free of dimension--5 operators. 
We have proposed a solution to this 
problem which  involves a texture of the pentaplet matrix together 
with certain constraints on the pentaplet couplings to matter.
 
 {\bf{\Large{Acknowledgments}}}
 
 M.G. wishes to acknowledge financial support from the TMR Network ``Beyond the Standard Model". K.T. acknowledges 
 support from the Programme $\Pi $ENE$\Delta$ 95. 
The work of J. R. was  supported in part by the EEC under the TMR contract  ERBFMRX-CT96-0090.

 \end{document}